\documentstyle{aipproc}
\begin{document}

\rightline{June 1996-- UCLA/96/TEP/20}

\newcommand{\gsim}{\mathrel{\raise.3ex\hbox{$>$\kern-.75em\lower1ex\hbox{
$\sim$}}}}

\title{Cosmology and Astroparticles\footnote{Talks given at the 
V Taller de Part\'{\i}culas 
y Campos (V-TPyC) and V Taller Latinoam.
de Fenomenolog\'{\i}a de las Interac. Fundam. (V-TLFIF),  Puebla, M\'exico,  10/30 - 11/3 1995.}}

\author{Graciela B. Gelmini}
\address{University of California, Los Angeles\\
Department of Physics and Astronomy\\
Los Angeles, California \\}
\maketitle

\begin{abstract}
These lectures are devoted to elementary particle physicists and 
assume the reader has very little or no knowledge  of cosmology and
 astrophysics.
After a brief historical introduction to the development of modern 
cosmology and astro-particles in which the Hot Big Bang model is defined, 
 the Robertson-Walker metric and the dynamics of the 
Friedmann-Robertson-Walker cosmology are discussed in  section 2.
 In section 3  the main observational features of the Universe are reviewed,
including a description of our neighbourhood, homogeneity and isotropy,
the cosmic background radiation, the expansion, the age and the matter content 
of the Universe. A brief account of the thermal history of the Universe follows
in section 4, and  relic abundances are discussed in section 5.
Section 6 is devoted to primordial nucleosynthesis, section 7
to structure formation in the Universe and section 8 to the
 possibility of detection of the dark matter in the halo of our galaxy.
In the relevant sections recent developments are included, such as
 several so called ``crisis" 
(the age crisis, the cluster baryon crisis and the nucleosynthesis crisis),
and the MACHO events that may constitute the first detection of dark matter
in the halo of our galaxy.

\end{abstract}

\subsection*{1. Historical Introduction}

Cosmology, from the greek $\kappa o\sigma\mu o\varsigma$, that means
``order'', is the study of the origin and evolution of the Universe.  The
initial point of
modern cosmology may be taken to be 1905 with the advent of Einstein's
Relativity Theory, extended to General Relativity in 1915.  The first
solutions to Einstein's equation for a homogeneous and isotropic but time
evolving Universe were
found by Friedman in 1929 and in 1935-1936 Robertson and Walker showed that
Friedman's metric could
be derived from homogeneity and isotropy alone, independently of General 
Relativity.  Friedman's
equations show that a Universe containing matter and radiation is expanding
(or contracting).  In
1929 Hubble found the first indication of  the recession of faraway galaxies
with a velocity $v$
proportional to its distance $d$, $v = H_o d$, where $H_o$ is called the Hubble
constant, the present value of the Hubble 
parameter (constant in space but not in time).
 Notice that the subscript zero  indicates a present value in cosmology.

Together with the belief
that we are not in a special place in the Universe, elevated to a principle
in modern cosmology,
the Cosmological Principle, the recession from us of distant galaxies means
that the Universe is
expanding (so that any two bodies not gravitationally bound to each other
are receding from each
other).
If the Universe was smaller in the past and it expanded
adiabaticlly, it was denser and hotter.  Gamow was
first in taking seriously
this idea. He proposed in 1946 that all the matter in the Universe was hot
and reached nuclear and larger densities in the past.  
This is the ``Hot Big Bang'' model \cite{Gamow}. 
 This model
implies a beginning of the Universe a time $t_o \simeq H_o^{-1}$ ago. 
  Hubble's estimate of $H_o$, the 
 Hubble constant, was around 5000 km/(sec Mpc), larger than the presently 
 known value of 40-100 km/(sec Mpc), what yielded a lifetime of the
Universe $t_o \simeq H_o^{-1} \simeq 4.5 \times 10^9$ yr, shorter than the already
known age of the Earth (1 pc = 1 parsec = 3.2615 light-yr).

 In part because the measured $t_o$ was
too short at the time, in part due to the philosophical difficulty of
accepting that the
Universe had a beginning, the competing ``Steady-State Universe'' model was
introduced by Bond, Gold
and Hoyle in 1948.  In this model the Universe is homogeneous also in time,
 it always looked as
it looks now.  Because in the Hot Big Bang model matter and radiation were
in equilibrium at early
times, a remanent of radiation with a black body spectrum is expected (that
would not be there in a
Steady-State Universe), as predicted by Alpher and Herman \cite{Alpher} in 1949. 
 The observation of this radiation by Penzias
and Wilson \cite{Penzias} (as explained by Dicke et al. 
\cite{Dicke}) in 1965, confirmed
the Hot Big Bang model.
It is remarkable that Alpher and Herman predicted the temperature of this
cosmic microwave
background radiation, CMBR, to be $5^o$K, quite close to its actual
temperature of $2.7^o$K.
Actually, the experimental confirmation of the Hot Big Bang already existed
when Gamow first
proposed it, in 1946.  Astronomers knew since 1941 that interstellar CN molecules
are excited in high
rotational levels, as seen through CN absorption lines in the spectrum of
stars, indicating that
those molecules are in a thermal bath at about $3^o$K.

Gamow's main interest in proposing the Hot Big Bang was the 
primordial  synthesis of nuclei.  It was soon
realized that the synthesis of heavy elements had to be done in stars,
 what lead to the study of
stellar evolution.  Only in 1966 the quantitative study of
primordial nucleosynthesis  started, the synthesis of 
${}^4$He at first and  that of  D, ${}^3$He and ${}^7$Li only later, 
in the early 70's.

This period, the early 70's,
can be considered as the beginning of astro-particles.  In 1967
(the year when
the standard electroweak model was proposed) Sakharov
noted the essential elements necessary
to generate the asymmetry between matter and antimatter in our Universe,
namely the generation of baryon number or
baryogenesis. One of them is the violation of baryon number,
  that encountered in Grand Unified Theories
(first proposed by Georgi and
Glashow in 1974) its most natural source.  In
the 70's cosmological
and astrophysical implications of  neutrinos were first studied,
leading to the battery of
tests that limit the number of neutrino  species, their mass,  cosmological
density, lifetime, decay modes
etc. Also, it was in 1975 when the first evidences for the dark matter were
discussed and the idea of inflation came about in 1980.
(For  references and more details about the content of this section
 so far, see e.g. \cite{Kolb})

 The last fifteen years can be
considered to be the golden age
of astro-particles, in which the connection of particle physics and cosmology has
lead to a number of important ideas.  Among them is the idea of
non-baryonic dark matter -see section 8-, that
is explored in simulations of galaxy formation confronted with
observational cosmological data and
in experimental dark matters searches, both with cryogenic  detectors and large
neutrino detectors,
supplemented by accelerator bounds. Other examples are the  models for baryogenesis and for
inflation, the interplay between
the properties of proposed particles and their cosmological and
astrophysical consequences, the introduction of particle physics experimental techniques
to study astrophysical
problems (such as in large ``neutrino telescopes'' or in the search for
MACHOS -see  section 8- in the halo of our
galaxy).  As an example of this last point, let us mention that experiments
designed to search for
the decay of protons predicted by Grand Unified Theories (IMB and Kamiokande)
observed for the first time neutrinos from a supernova, SN 1987A, 
 inaugurating with this
observation of  neutrinos from outside the solar system,
 what we can call neutrino-astronomy.

\subsection*{2. The Hot Big Bang Model}

\paragraph*{2.1 The Model --}

The Hot Big Bang (Hot BB), the standard model of cosmology,   establishes
that the Universe is homogeneous, isotropic (thus, it is a 
Friedmann-Robertson-Walker Universe) and  expanding from a
state of extremely high
temperature and density.  Thus the early Universe can be described as an
adiabaticlly expanding
classical gas of relativistic particles, namely radiation, in local thermal equilibrium at a
temperature $T$, that changes with time, $T(t)$.
 The lifetime of the Universe $t$ is counted from the
moment the expansion
started, taken to be $t = 0$.  

This model is based on General Relativity,
the Cosmological
Principle and three major empirical pieces of evidence, namely the Hubble
expansion, the cosmic
blackbody microwave background radiation (CMBR) and the relative cosmic
abundance of the light
elements (up to ${}^7$Li).  Because the model is based on General
Relativity, it is for sure not
valid in the realm of Quantum Gravity, $T > M_{\rm Plank} = 1.22 \times 10^{19}$
GeV and $t < 10^{-43}$ sec. 
 The Cosmological Principle postulates that we do not live in a
special place in the Universe,
by requiring that every comoving observer in the ``cosmic fluid" has the same
history.  The ``cosmic
fluid'' has as particles clusters of galaxies, and ``comoving'' in practice
means at rest with the
galaxies within a 100 Mpc radius (one parsec, 1 pc = 3.26 light-years =
3.09 $\times$ $10^{18}$
cm), as will become clear below (see section 3.2). 
 The Hubble parameter $H$ provides the
proportionality between the
velocity  $v$ of recession of faraway objects, and their relative distance $d$,
%
\begin{equation}
v = Hd.
\label{Hubble}
\end{equation}
The CMBR was produced at t$_{\rm rec}\simeq 3 \times 10^5$ y, the 
recombination epoch, when atoms became stable.  
It has a blackbody spectrum and it is remarkably isotropic. Finally, the
earliest available proof of the consistency of the Hot BB model is provided by the
cosmological abundance of ${}^4$He and of the trace elements D, ${}^3$He and ${}^7$Li.
 Their abundances, differing by several orders
of magnitude, are well accounted for in terms of nuclear reactions that occur
at $t_{\rm BBN} \simeq
10^{-2}-10^2$ sec, the Big Bang nucleosynthesis (BBN) epoch, when the photon
temperature was $T_{\rm BBN}
\simeq 10 - 0.1$ MeV (necessarily below the binding energy of the light nuclei). ~~~ Let us expand on these points.

\paragraph*{2.2 Friedman-Robertson-Walker Models}

Homogeneity and isotropy restrict  the space-time metric to be of the
Friedmann-Robertson-Walker form,
%
\begin{equation}
ds^2 = dt^2 - a^2(t) \left [ {dr^2\over 1-kr^2} + r^2 (d\theta^2 + \sin^2
\theta d\phi^2)\right]~.
\label{metric}
\end{equation}
This metric depends on a global scale factor $a(t)$ that describes the overall
expansion of the Universe, i.e. a physical
distance $\lambda$ increases with the Hubble expansion as $\lambda = a(t) \lambda_c$,
 where $\lambda_c$ is the constant distance measured in a
comoving reference frame (a frame that follows the Hubble expansion, like a grid
painted on an expanding balloon), whose coordinates are $r, \theta$ and $\phi$.  So
$\lambda_c, r,\theta, \phi$ do not change with the expansion of the Universe.
With the usual choice of $a_o$=1 for the present value of the scale factor,
 $\lambda_c$ corresponds to the present physical distance. The time coordinate
$t$ in Eq. (\ref{metric})  is the proper time measured by a
comoving observer, an observer 
(located at fixed $r, \theta$ and $\phi$)
 that moves with the local Hubble flow. $k$ is the curvature parameter.
 The spatial curvature, ${}^3K(t) \equiv k/a(t)^2$, can be either
positive (corresponding to a closed Universe with the geometry of a
3-sphere), or  zero
 (corresponding to
a flat open Universe), or negative (corresponding to an open Universe with
the geometry of a
3-hyperboloid) depending on the value of the curvature factor.  Through
rescaling of the coordinates, $k$
 can be chosen to be $+1,$ 0 or $-1$ respectively.

This metric yields the Hubble expansion if the Hubble parameter  $H(t)$, defined as $H(t) \equiv \dot a(t)/a(t)$, 
is positive, since taking $t, \theta$, and $\phi$ constant, $ds = d D(t)$ where
$D(t)$ is the proper
distance, $D(t) \equiv a(t) \int_o^r dr(1-kr^2)^{-1/2}$, so that the
proper velocity $V \equiv
dD(t)/dt$ follows Hubble's law, $V \equiv HD$.  However $V$ and $D$ are not
operational quantities
(those that can be measured) such as the 
velocity of recession $v$ and the
luminosity distance
$d_L$ (see, for example,\cite{Kolb}).  $v$ is measured through the red-shift $z$ of the 
observed light (of wavelength $\lambda_o$)
with respect of the emitted light (of wavelength $\lambda_e$),
$1+ z \equiv {\lambda_o / \lambda_e}  = {a(t_o) / a(t_e)} \simeq {v/ c}$,
where $c$ is the velocity of light. This is the non-relativistic expression, 
only valid for $z\ll 1$, otherwise $(v/c) = [(1+z)^2 -1]
[(1+z)^2 + 1]^{-1}$. $d_L$ is 
measured through the absolute luminosity $\cal L$ 
 of the emitting body 
(whose difficult determination is the origin of the uncertainties in the
measurements of $H$)
and the measured flux $\cal F$, $d_L^2$ $\equiv$ $\cal L$/ $4\pi \cal F$.  Thus 
Hubble's law
%
\begin{equation}
H_od_L \simeq z + {1\over 2} (1-q_o) z^2 + \cdots
\label{Hubblequad}
\end{equation}
reduces to  Eq.(1) with
$d = d_L$, for $v\ll c$.
The deceleration parameter $q_o \equiv - (\ddot a a/\dot a^2)_o$, if not zero, becomes
important at moderate $z$, where yet poorly known changes in the sources due to their 
evolution also become important. For this reason $q_o$ is only known to be at most of
 order one. 
 $H_o$, the Hubble constant, the present value of the Hubble
parameter $H$,  
is at present observationally determined within a factor of 2 to be
%
\begin{equation}
H_o = 100~ h {{\rm km}\over { \rm sec~ Mpc}}~,
\label{H0}
\end{equation}
with  $h$= 0.4 -1.  Through Hubble's equation, recession velocities or 
red-shifts are translated into distances,
%
\begin{equation}
d_L \simeq 10^{-2}~ {cz \over{\rm km/sec}} ~h^{-1} {\rm Mpc}=
 3000 ~z~h^{-1}{ \rm  Mpc}~,
\label{dL}
\end{equation}
for $z < 1$. 
Notice that the units of distance are ($h^{-1} {\rm Mpc}$) in this case.

\paragraph*{2.3 Evolution of the Universe --}

The dynamics of the global scale factor $a(t)$ is determined by the content
 of the Universe (in matter,
radiation and vacuum energy) through the Einstein field equations,
$R_{\mu\nu} = {1 \over 2} R
g_{\mu\nu} = 8\pi G T_{\mu\nu} - \Lambda g_{\mu\nu}$. Spatial
 homogeneity and
isotropy require the total
stress-energy tensor $T_{\mu\nu} $ to be diagonal and with equal space
components.  The simplest
realization corresponds to a perfect fluid $T_{\mu\nu} = $ diag $(\rho(t),
- p(t), -p(t), -p(t))$
where $\rho(t)$ and $p(t)$ are the energy density and the pressure
respectively.  $\Lambda$, the
cosmological constant introduced by Einstein in 1917 in order to obtain a
static Universe (idea
rejected experimentally by the discovery of the Hubble expansion in 1929),
is associated in modern
terms with vacuum energy. It can be incorporated into $T_{\mu\nu}$ by adding
$\rho_{\rm vac} =
\Lambda/8\pi G$ and $p_{\rm vac} = - \rho_{\rm vac}$ to the $\rho$ and $p$
of radiation and
matter to obtain  $\rho_{\rm total} = \rho + \rho_{\rm vac}$ and $p_{\rm total} = p +
p_{\rm vac}$.  With
this $T_{\mu\nu}$ there are only two independent Einstein equations, 
from the 00 and $ii$
components respectively.  From
$\mu = 0$, $\nu = 0$
we obtain the Friedmann equation,
%
\begin{equation}
H^2 = \left ({\dot a\over a} \right )^2 = {8\pi G\over 3} \rho - {k\over
a^2} + {\Lambda \over 3}~,
\label{E1}
\end{equation}
whose Newtonian meaning of energy conservation for a unit mass test
particle can be seen (for
$\Lambda = 0$) by writing  it as $\dot a^2/2 - GM/a = - k/2$ for $M = (4/3)$
$\pi a^3\rho$, where the
total energy is $-k/2$.  Combining the $00$ and $ii$ components one gets
%
\begin{equation}
{\ddot a\over a} = {-4\pi G\over 3} (\rho + 3p) + {\Lambda\over 3}~,
\label{E2}
\end{equation}
that for matter (gas, dust), for which $ p = 0$ (and $\Lambda = 0)$, has the 
meaning of acceleration equal
force over mass, $\ddot{a}= - GM/a^2$.

In order to determine the three unknown function of time in these
equations, $a(t)$, $\rho_{\rm
total} (t)$ and $p_{\rm total}(t)$ we need a third independent equation.
This is the equation of
state, $p = p(\rho)$ provided by thermodynamics.  For only relativistic
particles and radiation,
$p_R = \rho_R c ^{2}/ 3$, while for only matter (gas, dust) $p_M = 0$
(for a gas $p_M = n k T \sim \rho_M
v^2 \ll \rho c^2$, where $n$ is the number density and $k$ is the Boltzmann
constant).  It is
convenient to introduce the energy conservation equation, 
$T^{\mu\nu}_{,\nu} = 0$ for $\mu = 0$,
that is not independent of Eqs. (\ref{E1}) and  (\ref{E2}), 
$\dot \rho = - 3(\dot a/a) (\rho+ p)$.  Using the
equations of state in this equation, one obtains $\rho$ as a function of
$a$, $\rho_R \sim a^{-4}$
and $\rho_M \sim a^{-3}$, and using these in Eq. (\ref{E1}) one obtains,
%
\begin{equation}
a_R \sim t^{1/2},~~ H_R = {1\over 2t} ,~~ a_M \sim t^{2/3}, ~~ H_M =
{2\over 3t}~.
\label{aH}
\end{equation}
Thus, given a certain value of $H_o$, a Universe radiation dominated for
most of its life is
younger than a matter dominated one.
The relations in   Eq. (\ref{aH}) can be easily obtained by
using $a \sim T^{-1}$
(this is the entropy conservation condition, derived from the energy
conservation equation in
thermal equilibrium).  Then, we know that $t^{-1} \sim H \sim  \sqrt\rho$
(from the Friedmann
Eq. (\ref{E1})) and we know that $\rho_R \sim T^4$ and $\rho_M \sim m/a^3
\sim T^3$, therefore $t\sim
T^{-2}\sim a^2$ for radiation and $t \sim T^{-3/2} \sim a^{3/2}$. 

With only
vacuum, namely with $p = \rho = 0$ and  $\Lambda > 0$ constant, 
neglecting the curvature term we obtain from Eq. (\ref{E1})  an
exponential expansion of the scale
factor $a \sim e^{HT}$ with $H = \Lambda/\sqrt3$.
 Notice that 
with an exponential expansion the curvature term
becomes negligibly very fast, $k/a^2 \ll {\Lambda / 3}$.  
This is precisely why
a period of vacuum dominated expansion in the early Universe, called
``inflation", has been proposed
to explain the flatness (see section 3.7) of the Universe. 

 Inflation would happen during a
period of supercooling of the Universe 
(so that the density of radiation and matter,
proportional to $T^\alpha$,
$\alpha = 4, 3$ become negligible) while the Universe is rolling down the
almost constant potential  $V(\phi) \simeq V_o$   of a scalar field, 
 the ``inflaton" $\phi$, so that $\rho_{\rm vac} \simeq
V_o \gg \rho_R, \rho_M$.  At the end of inflation it is necessary to
convert $V_o$ into radiation
and matter, through the decay of the $\phi$ field, in a process called
``reheating",  that generates
the Universe as we know it, still with a negligible curvature. 
We can rewrite Eq. (\ref{E1}) in the
form
%
\begin{equation}
{k\over H^2a^2} = {\rho_{\rm total}\over (3H^2/8\pi G)} -1 
\equiv {\rho_{\rm total}\over \rho_c} -1\equiv \Omega - 1
\label{Omega}
\end{equation}
where $\Omega$ is the density in units of the critical density,
$\rho_c \equiv$ $ {3H^2/ (8\pi G)}$ $= 10.5 ~h^2 {(keV/ cm^3)} $ $= 1.88~
10^{-29}  h^2 {(g / cm^3)}$. 
%
\begin{figure}[b!] 
\vspace{7.5 truecm}
\vspace{10pt}
\caption{Evolution of the cosmic scale factor $a(t)$ in different 
FRW  models.}
\label{fig1}
\end{figure}
Notice that spatially open, flat and closed Universes (with $k < 0,$ $k =
0$ and $k > 0$
respectively) correspond to $\Omega < 1$, $\Omega = 1$ and $\Omega > 1$
respectively.  So,  a
long enough period of inflation produces $a^{-2} \rightarrow 0$, thus
$1 - \Omega^{-1} \sim (k/a^2\rho_{\rm total})
\simeq 0$ (since
$\rho_{\rm total} = \rho_{\rm vac} $ is constant). That is, $\Omega = 1$ after a long
enough inflation. 

Notice that only for $\Lambda = 0$ there is a correspondence between
spatial geometry and the
future evolution of the Universe, at $t \rightarrow \infty$. 
 Let us  call here $\Omega_o$ the density of matter and radiation,
 $\Omega_o = \Omega_R + \Omega_M$, so that the total density is 
$\Omega= \Omega_o + \Omega_\Lambda$.
 Because for
matter and radiation $\rho
\sim a^{-\alpha}$ with $\alpha = 3$ or 4, 
for large values of the scale factor $a$ the curvature
factor $k/a^2$
dominates the $r.h.s$ of Eq. (\ref{E1}). 
In this case, an open or flat Universe, 
i.e. with $\Omega = \Omega_o \leq 1$ $(k \leq 0)$, 
expands forever and a closed Universe, with $\Omega = \Omega_o > 1$ $(k > 0)$
recollapses.
 A cosmological  constant $\Lambda$ dominates instead the $r.h.s.$ of
Eq. (\ref{E1}) for large
values of $a$, and  for $\Lambda > 0$ the Universe expands exponentially 
forever at large
times (even if $\Omega \geq 1$), as can be seen in Fig. 1 
(taken from \cite{Frieman}).

 Notice that for matter or radiation $\rho + 3p > 0$, what,
 through the Eq. (\ref{E2}) 
(with $\Lambda = 0)$, implies that $\ddot{a}< 0$, so the
expansion of the Universe is slowed down by the gravitational pulling of
the mass content of the
Universe, and consequently  $t < H^{-1}$ (since $t = H^{-1}$ corresponds to
$\ddot{a}= 0)$. On the other hand, 
 for a vacuum dominated Universe the
pressure is negative,
$p_{\rm vac} = - \rho_{\rm vac}$ and Eq. (\ref{E2}) shows that gravity is
repulsive, $\ddot{a} = \Lambda a/3 > 0$ and consequently $t > H^{-1}$. 
 Thus,  a non-zero cosmological constant
yields larger values of $t_o$ for a given
$H_o$ (see Fig. 1) than matter and radiation alone. We see, therefore,
how a non-zero $\Lambda$ can resolve the present
possible conflict between a
large value of $H_o$ and a large value of $t_o$, the ``age crisis" we will
mention later (see section 3.6). 
(For references and more details about this section see e.g. \cite{Kolb})

\subsection*{3. Observational Features Of The Universe}

\paragraph*{3.1 Topography of Our Neighbourhood --}

Galaxies are the building blocks of our Universe.  Normal galaxies have
masses of $10^8$ to
$10^{12}$ $M_\odot$ (where $M_\odot$ is the solar mass), while dwarf galaxies
 have $10^6$ to
$10^7M_\odot$ (same mass as that of globular clusters, old systems of stars
spherically
distributed around the center of galaxies like ours).  We live in a spiral
galaxy, the Milky Way,  with a thin disk of a few 100 pc of thickness
and  approximately 12 kpc of radius. The Sun is in one of its arms, 
at about 8 kpc from the center. 

 Galaxies form gravitationally bound systems:  binary systems, 
groups (that are systems of a few to 10 galaxies, with a typical size of 
1 Mpc), clusters (systems with hundreds of
galaxies extending from a few to 10 Mpc) and superclusters (with thousands of
galaxies in a radius of 10-50 Mpc). 
 We live in a group,
 the Local Group, together with another large spiral galaxy, Andromeda,
  at roughly 700
kpc of the Milky Way, and several small galaxies.
 
At the scale of superclusters, i.e. scales of 20-100 Mpc, the Universe 
is better described in terms of ``walls", ``filaments" and ``voids" \cite{CfA}. 
 There are few galaxy maps at even larger scales, i.e. scales larger than 100
Mpc, and they show smoothness at those scales.  Present maps only cover a
small fraction of
our visible Universe, whose radius is 
$ct_o \simeq$ $10^{10}$ light yr $\simeq 3000$  Mpc.

While  Hubble's law and red-shift measurements are the only tools to 
obtain the distance of very faraway objects, the distance to nearer
ones can be estimated separately from their red-shift, what allows to measure
``peculiar" velocities $v_{\rm pec}$, i.e. velocity
components  due to gravitational acceleration that add to the local
velocity due to the Hubble
flow $H_o d$ to give the total velocity $v = v_{\rm pec} + H_o d$.
In this way it has been determined that large regions of about 100  Mpc of
our local neighbourhood are moving
towards a very massive system, called the Great Attractor, with 
 large peculiar velocities of about 600 km/sec \cite{samurai} \cite{Deckel}.

\paragraph*{3.2  Homogeneity and Isotropy --}

Although these are assumptions incorporated into the BB model by the
Cosmological Principle, there
are strong observational foundations of the homogeneity and isotropy of the
Universe at 
scales larger than 100 Mpc.  

As we have just seen,
the Universe looks lumpy and  large regions move with large
peculiar velocities at  scales $\lambda < 100$ Mpc, 
however the Universe looks smooth at larger
scales and we also know it was smooth in the past. 
 There are few 3-dimensional galaxy maps 
that reach the necessary large
scales to test homogeneity, containing  in total only of the order of 
$10^4$ galaxies,
 but by the end of the
century $10^6$ will be
mapped. These are red-shift surveys where distances are obtained
from $z$ though Hubble's law.
Angular  photometric galaxy surveys instead,  provide 2-dimensional
information for several
$10^6$ galaxies.  Both types of galaxy maps show smoothness at 
$\lambda > 100$ Mpc, in the case of
2-dimensional maps by showing that angular correlations decrease steeply
with increasing angular
distance in the sky (see for example \cite{Frieman} for references). 
 But the best evidence of homogeneity and isotropy is
provided by the isotropy
of the CMBR, for which measurements of the temperature in different
regions of the sky show very small anisotropies, smaller than $10^{-4}$,
i.e. the r.m.s. temperature fluctuation measured between two points 
in the sky separated by an angle $\theta$ is
 $ ( {\delta T / T} )_{\theta} \lesssim 10^{-5}$ \cite{GFS}
  \cite{White}.
This shows both the smoothness of the Universe at the moment of emission of
the radiation and the isotropy of the expansion since then,
 because the present
measured wavelength of the CMBR photons now is
$\lambda_{\rm now} = a(t)  \lambda_{\rm e}$, where $\lambda_{\rm e}$
is the wavelength at emission,
and any space dependence of the scale factor
$a(t)$ would produce anisotropies. The  CMBR photons are emitted 
from  the last
scattering surface, that is the Universe at recombination
($t \simeq 3 \times 10^5$yr, $T \simeq 0.3$ eV and
$z \simeq 1100$). 

\paragraph*{3.3.  Cosmic Microwave Background Radiation --}

The best blackbody in the Universe is the Universe.  The COBE (Cosmic
Background Explorer)
Satellite measured  the CMBR spectrum  in 1992 for wavelengths between 
0.05 and 1 cm, and found a blackbody
with temperature $T_o = 2.726 \pm 0.010 {}^o$K, with deviations of less
than 0.03\% \cite{Mather}.  This provides
fundamental evidence for a Hot BB. As we just mentioned, 
the CMBR was produced at the
recombination epoch, when atoms
became stable and replaced ions and electrons in a plasma as the
constituents of the Universe.  At
this moment the mean free path of photons, that was very short in the 
preceding plasma, became very long, so
that the photons that last scattered then can reach us.  Knowing so well
the CMBR temperature, we
know with great accuracy the number and energy density of the CMBR  photons,
that  are the
most abundant in the Universe by several orders of magnitude (see e.g.
\cite{Kolb} p. 143),
$n_\gamma = {2\zeta (3) T_o^3 / \pi^2} = {411/ cm^3}$, $\rho_\gamma = 
{\pi^2 T_o^4 /15} =
4.71 \times 10^{-34} {(g/ cm^3)}$
Using  $\rho_c$ in Eq. (\ref{Omega}), 
we see that $\Omega_{ \gamma} \simeq 10^{-5}$.

Anisotropies in the CMBR are expected due to our  motion with respect to
the CMBR rest frame, and due
to the density inhomogeneities that triggered structure formation in the
Universe.  In fact COBE
measured a dipole anisotropy corresponding to a velocity of 627 $\pm$ 22 km/sec 
 of our Local Group of
galaxies with respect to the
CMBR rest frame  (COBE even saw the rotation
of the Earth around the
Sun!), and measured anisotropies $(\delta T / T)_{\theta}$ for angles 
$\theta = 7^o$ to $90^o$.  At $90^o$ COBE measured
a quadrupole anisotropy of 
$(\delta T/T)_{90^o} \simeq 0.5 \times 10^{-5}$ \cite{GFS}.
  Results from other experiments in
balloons are available now at smaller angles, $\theta = 0.5^o$ to $90^o$
and the results show $\delta
T/T \lesssim 10^{-5}$ after subtracting the dipole.  

The horizon size at
recombination, $c t_{\rm rec}$,
corresponds to an angle \break $\theta_H \simeq 1^o (\Omega)^{-1/2}$ in the
present sky. Larger angles correspond to causally disconnected regions at 
the time of emission of the CMBR
photons.  Thus the smallness of the anisotropies at angles $\theta \lesssim
1^o$ indicates that
the Universe was smooth at recombination, and at larger angles, $\theta \gsim
1^o$ it indicates that
the Universe is smooth on very large scales.  The temperature autocorrelation
function, $C(\theta)  =~ <
(\delta T (\alpha)/T) ~~(\delta T (\alpha + \theta)/T) >_\alpha~$,
 where $\delta T/T = (T - \bar T)/ \bar T$ are temperature fluctuations
 with respect to the average temperature
 $\bar T = 2.726 {}^o$K,  is computed by 
measuring $\delta T/T$ at some position
$\alpha$ in the sky ($\alpha$ is given in spherical harmonics by two angles), 
multiplying it with $\delta T/T$ at another position 
 $(\alpha + \theta)$  separated from the first one by an angle 
 $\theta$ and
averaging over all positions.  
\begin{figure}[b!] 
\vspace{7.5 truecm}
\vspace{10pt}
\caption{CMBR anisotropy power spectra  predicted by four models
(lines) of structure formation (see section 7). 
The band shows the expected experimental error in future
satellite experiments.}
\label{fig2}
\end{figure}
The expansion of $C(\theta)$ in Legendre polynomials defines the
multiple moments $C_{\ell}$,
$C(\theta) = (4\pi)^{-1} \sum_{\ell} (2\ell + 1) C_{\ell}P_\ell(\cos\theta)$,
so that the
anisotropy at a certain angle $\theta$ is 
%
\begin{equation}
\left ( {\delta T \over T} \right )_{\theta} \simeq
\sqrt{\ell(\ell+1)C_\ell}~, ~~~~~~\ell \simeq (200^o/\theta)~.
\label{deltaT}
\end{equation}
 We see that $\ell \geq 200$ correspond to 
scales within the horizon at decoupling,
where matter can move due to density perturbations and where the photons
pick up additional energy due to the scattering from moving matter
(mainly Thompson scattering on electrons).
This gives origin to Doppler (or acoustic) peaks in the spectrum of multiple
moments.
The position of the first peak should happen at the scale of the horizon
$\ell_{\rm peak}
\simeq 200 (\Omega)^{-1}$, as shown in Fig. 2  \cite{Scott}
for different dark matter
models (see section 7).  Thus, in the future, the position of the first
peak will determine
$\Omega$ within a $10\%$.  The height of the peaks depends on the density
in baryons, $\Omega_B$ and
the location and relative height of the peaks will allow to discriminate
different models of galaxy
formation, once all $C_\ell$ are measured. Actually only the $C_\ell$ for $\ell
< O(10^3)$ are relevant, because larger
values of $\ell$ correspond to scales $\theta < 8^o$ that are inside the
thickness of the last
scattering surface, so the information at those small scales is 
smeared out \cite{White}.

\paragraph*{3.4  The Expansion of the Universe --}

Hubble's law in Eq.(\ref{Hubble}) is valid at $z \ll 1$.  
At $z \gsim 1$ the law
  becomes quadratic (see Eq. (\ref{Hubblequad})) and
the deceleration parameter $q_o$  measures
the rate at which the
gravitational attraction is slowing down the expansion.  Actually $q_o =
\Omega_o/2 -
\Omega_\Lambda$ (for matter dominated $\Omega_o$), so for a large 
cosmological constant component 
$\Omega_\Lambda$,
$q_o$ could be negative
and the expansion could be accelerated (recall that gravitation is repulsive
in a vacuum dominated Universe).  Not much can be said
observationally about $q_o$ yet.  So
let us concentrate on the Hubble constant $H_o$.  

Redshifts can be measured accurately.  This allows to determine the
recession velocities $v$. In general they have two components $v= H_o d +
v_{\rm pec}$. Thus recession velocities
 can  be attributed solely to the Hubble flow 
at distances large enough for the peculiar radial velocities
$v_{\rm pec}$, the velocity components  due to
gravitational acceleration, to be negligible, 
$H_o d \simeq v \gg v_{\rm pec} $, what happens at $d \gg h^{-1}$ Mpc,
say distances larger than 50 Mpc,
because $v_{\rm pec} \simeq O (100$ km/sec). 

In order to measure $H_o = v/d$ (Eq. (\ref{Hubble})), 
distances and velocities must be determined 
separately.
   The dificulty in the measurement of $H_o$ reside in  determining $d$. 
Different techniques to measure
distances lead to the spread of roughly a factor of two in results for $h$
(the Hubble constant in units of 100 (km/sec Mpc) see Eq. (\ref{H0})).
Most methods give large values of $h$, $h \simeq 0.8$. For example, a
recently publicized result of
the Hubble Space Telescope is $h = 0.8 \pm 0.17$ \cite{Freedman}. 
 This determination is based on the
observation of twenty Cepheids in the Virgo cluster. Cepheids are variable
stars, whose period of
variability is associated with their intrinsic luminosity, thus, the
apparent luminosity tells the
distance to the star.  Some determinations give smaller values of $h$.
For example, using type
$Ia$ supernovae a determination of $h = 0.5 \pm 0.1$ was made 
(see e.g. \cite{Deckel} for references).

The problem with a large value of $h$, $h \gsim 0.65$, is that it yields a
too young Universe, unless $\Lambda \ne 0$ or $\Omega_o < 1$,
 namely unless we live in a Universe with positive cosmological
constant, or open, or a combination of the two.  This goes against aesthetic
beliefs in a flat, matter dominated Universe, giving origin to what 
some call an ``age crisis" (see section 3.6 below).

\paragraph*{3.5  The Age of the Universe --}

There are three main methods to determine $t_o$.  The technique of nuclear
cosmochronology, uses
the relative abundance of radioactive isotopes at present and at
production (through rapid neutron
capture processes, $r$-processes, most probably in supernovae).  Comparing
the two gives $t_o = (10-20) \times 10^9$  yr \cite{Cowan}.

White dwarfs are the last stage of low-mass stars. Since they are faint they
can only be seen near the solar system. There is a drop in the
number of white dwarfs with luminosity smaller than $3 \times 10^{-5}$ 
of the solar luminosity.
Combining cooling models with the assumption that this cut indicates that
there are no older white
dwarfs that had the time to become fainter, gives the age of the disk of
our galaxy.  This method
gives $t_o = (9-10)\times 10^9$ yr \cite{Winget}.

Globular clusters are old systems of a few-million stars that formed 
all at the same time.  The
evolution of stars depends on their initial mass, the more massive have
shorter lifes, and
finish burning Hydrogen earlier.  The age of globular cluster is determined
by measuring the absolute
luminosity of the stars that are finishing burning H, using well known stellar
evolution theory.  The
lifetimes obtained in this way are $t_o
\simeq (13-15) \times 10^9$  yr and it is very difficult to get to 
$t_o < 11-12 \times 10^9$  yr \cite{Renzini}. Thus, $t_o \gsim 13 $ Gyr is taken
at present to be  a reasonable lower bound of the age
of the Universe and  $t_o \gsim 10$ Gyr is taken to be an absolute lower bound, 
a minimum age already uncomfortable to accommodate globular clusters. 

\paragraph*{3.6  Age Crisis? --}

The parameters $t_o$, $H_o$ and $\Omega$ are not independent, actually
 $H_o t_o = 1.06$ $(h/0.80) (t_o/13 \times 10^9 $ yr) = $f_i (\Omega)$, where 
 $f_i$ is a function of $\Omega$ that depends on the
content of the Universe (see e.g. \cite{Kolb}).  For a  flat $\Omega_o = 1$ Universe  matter 
dominated by the dark  matter  (see section 3.7) with $\Lambda = 0$
(until recently  the model preferred by most cosmologists due to its
simplicity and aesthetic appeal), one has
$H_o t_o = 2/3$ what implies that $t_o  \gsim 13$ Gyr requires $h \lesssim 0.50$
and $t_o \gsim 10$ Gyr requires $h\lesssim 0.65$.  
As mentioned above, in section 3.4, these values of $H_o$ are lower than
 most present determinations (but not all).  If $h$ is
actually larger than 0.65, then we live in a Universe with a non-zero
cosmological constant, or open, or both.

If our Universe is
spatially flat, namely $\Omega = \Omega_o + \Omega_\Lambda = 1$, the frequency
of gravitational lensing of quasars by nearer galaxies gives a bound of
$\Omega_\Lambda \lesssim 0.7$ \cite{Fukugita}.  
The number of lensings depends on $\Lambda$, because with $\Lambda >0$ the 
distance to a
quasar of a given redshift is larger than for $\Lambda = 0$.  With $\Omega
= 1$ and $\Omega_\Lambda
\leq 0.7$ one gets $H_o t_o < 0.96$.  In this case $t_o >$ 13 (10) Gyr
requires $h <$ 0.72 (0.94).
 On the other hand if $\Lambda = 0$ and $h > 0.65$, we need 
$\Omega (= \Omega_o) < 1$, namely we live in an open
Universe.  For a matter dominated Universe $H_o t_o$ increases for
decreasing $\Omega_o$, from 2/3 for $\Omega_o = 1$  to 1 for $\Omega_o = 0$
(namely for an empty Universe $t_o = H_o ^{-1}$,
since $\ddot a = 0$).
Dark matter measurements indicate that there is at least $\Omega_o = 0.1$ in
matter, what means $H_o t_o  \lesssim 0.9$. In this case
(with $\Lambda = 0$) $t_o > 13 (10)$ Gyr
 requires $h <$ 0.68 (0.88).  
Of course a
combination of $\Lambda \ne 0$ with $\Omega < 1$ would also be possible,
but even less appealing.

Why does $\Lambda\ne 0$ or $\Omega < 1$ seem so unappealing?  There is a
problem of fine tuning in both cases.  A cosmological constant corresponds 
to a non-zero vacuum energy.  We cannot explain why the vacuum energy 
should be zero at present, but it is even less appealing to explain why
 if non-zero it is so small with
respect to characteristic values of the potential energy of scalar fields
in particle physics.  Take
as an example the usual Higgs potential responsible for the spontaneous
breaking of the electroweak symmetry.  
Take $\phi$ to be the usual Higgs doublet, then the potential energy density
is $V = \lambda (|\phi|^2 - v^2)^2$ and the
hight of this potential is $V_o = V (\phi = 0) = \lambda v^4 = \lambda (100$
GeV$)^4$.  In  units of GeV$^4$
the critical density is $\rho_c = 10^{-46} $ GeV$^4$, so $\rho_{\rm vac}=
\rho_{\Lambda} \lesssim \rho_c = 10^{-54}
\lambda^{-1}V_o $! 

With $\Omega \ne 1$ there is an equally important fine tuning to take into
account.  If $\Omega\ne 1$, $\Omega$ increasingly separates from 1 as $t$
increases in a Universe
radiation or matter dominated.  Dividing Eq. (\ref{Omega}) by $\Omega$, 
we find $1-\Omega^{-1} = 3 K
(8\pi G a^2 \rho)^{-1} \equiv \chi (t)$ and $\chi(t) \sim a$ for matter
$\sim a^2$ for radiation (see Eq.(8)). Thus, in order to
have $\Omega = O(1)$ (but
$\ne 1$), $\Omega$ should have differed from 1 in the past by an
extremely small amount, at most
$\chi(t) \simeq 10^{-10}$ at the nucleosynthesis epoch
($T\simeq$ 1 MeV) or
$10^{-32}$ at the Planck time $(T\simeq 10^{19}$GeV).  These
small initial values of $\chi(t)$  are unappealing.

\paragraph*{3.7  Matter Content of the Universe --}

The luminous mass, namely the matter associated with typical stellar
populations,  accounts for $\Omega_{\rm Lum} \simeq 0.003 h^{-1}$ \cite{Trimble}. 
It is determined through luminosity measurements knowing  the
average mass over luminosity ratio. This is, for example,
 5 times the solar ratio in the solar neighborhood  
and 12$h$ times the solar ratio in the core of elliptical galaxies.

However there is much more  than the luminous mass in the Universe.
It is by now a well established fact that on all scales larger than about
10 kpc there is a discrepancy between the amount and distribution of the 
luminous mass, and the mass
detected through its gravitational effects.  The gravitationally dominant
mass component of the Universe is ``dark", i.e. it is not seen either in 
emission or absorption of any type of
electromagnetic radiation.  This is called dark matter (DM).  The most
robust evidence for the DM
is given by the rotation curves $v(r)$ of spiral galaxies, i.e. the orbital
velocities of stars and
gas clouds orbiting a spiral galaxy at a distance  $\it{r}$ from its
center.  These velocities
remain constant outside the region where the light falls exponentially off.
Since $v(r)^2 =
GM(r)/{r}$, the flatness of the rotation curves $v(r)$ beyond a few kpc
from the center means
that the mass contained within a radius $r$ grows linearly with $r$, $M(r)
\sim r$.  This indicates
the existence of a quasi-spherical DM halo around each galaxy whose
density falls off as $r^{-2}$
outside a core radius of a few kpc.  These DM haloes amount to
$\Omega_{\rm DM} = 0.02 - 0.05$ \cite{Faber}\cite{Trimble}.
The first indications of the necessity of DM came from the study of the
Coma cluster by Zwicky in 1930.
Applying the virial theorem to clusters of galaxies the total amount of
matter in the cluster
can be inferred from the characteristic velocity and average separation of
its galaxies.  The DM
measured in this manner is $\Omega_{\rm DM} \simeq $ 0.1 - 0.2 \cite{Trimble}.  
A more recent method uses the
gravitational lensing of very far galaxies by clusters, whose total mass can
be reconstructed from
the deformation of the lensed galaxies into arcs and arclets.  This method
also yields
$\Omega_{\rm DM}\simeq $ 0.1 - 0.2 \cite{Lynds}.  Peculiar velocities are due to
gravitational accelerations due to
 masses nearby.  Large scale peculiar velocity flows, at scales
of 50 Mpc and larger,
can then be used to infer total mass distributions, what yields estimates
in the range $\Omega_{\rm DM}$
$\simeq$ 0.2 - 1 \cite{Deckel}.

With respect to the density in baryons (namely nuclei) $\Omega_B$, Big Bang
nucleosynthesis (BBNS)
predicts the density of light elements in agreement with observations only
for $\Omega_B h^2 \simeq 0.01
- 0.02$ (see more about nucleosynthesis in section 6),
 what for $h = 0.4 - 1$ means $0.01 \lesssim \Omega_B
\lesssim 0.13$.  This means that all the DM in the halo of galaxies could be
baryonic (in the form of
macroscopic objects, such as failed stars or primordial black holes
 for example), but also may
indicate that some of the
baryons are dark $(\Omega_B > \Omega_{\rm Lum})$ if $h < 1$.

The BBNS estimate of $\Omega_B$
combined with the measurement of a ``large'' amount of gas in rich clusters
of galaxies has led to
what some call ``the x-ray cluster baryon crisis''\cite{White2} \cite{Steigman}.
  In fact, the gas in
the central part (until a
radius of about $1 h^{-1}$ Mpc) of the cores of rich clusters is seen
through the x-rays emitted by
the gas in hydrostatic equilibrium and it has been recently seen that the
fraction of the total mass in gas in those regions is 
large, f = ($M_{gas}/M_{total}$) $\simeq (0.05 -
0.10)h^{-3/2}$.  Since the clusters
in question are large clusters, one can think that this ratio is
representative of the baryon
fraction in the Universe $\rho_B/\rho_{M} = \Omega_B/\Omega_{\rm DM}$.  If
so, $\Omega_{\rm DM} =
\Omega_Bf^{-1}$ and using the measured large value of f and the BBNS upper
bound  of 0.02$h^{-2}$ on $\Omega_B$ one gets an upper limit of
$\Omega \leq (0.2 - 0.4)h^{-1/2}$ \cite{White2}. If $h > 0.16$ 
(as all measurements confirm, see section 3.4) and
 the bound on $\Omega_B$ from
BBNS is correct, this would
mean that either we live in an open Universe  (if $\Lambda = 0$)  or 
in a Universe with $\Lambda \not= 0$ (if we want it flat
we need $\Omega_{\rm DM} + \Omega_{\Lambda} = 1$) or  
both.

\subsection*{4. Thermal History Of The Universe}

As we have seen in sections 3.3 and 3.7, the present energy in radiation
$\Omega_{\gamma} \simeq
10^{-5}$, is much smaller than that of matter $\Omega_0 \simeq
\Omega_{\rm DM} \simeq 0.1-1$.  This
still holds true even when adding the contribution of relativistic
neutrinos, with which
$\Omega_R (T_o) = 4 \times 10^{-5} h^{-2}$ (g$_*$/3.36), where g$_*$ are
effective relativistic
degrees of freedom and for photons and three relativistic neutrino species,
g$_*$ = 3.36 (see section 5).
Due to the
different evolution with temperature of the density of matter and
radiation, $\rho_R \sim T^4$ and
$\rho_M \sim T^3$, the radiation
was dominant in the past at $T > T_{eq}$, where $T_{eq}$ is the temperature
at matter-radiation equality,
$\rho_R(T_{eq}) = \rho_M(T_{eq})$,
$T_{eq} \simeq 5.8 {\rm eV}~ \Omega_o H_o^2 (3.36/g_*(T_{eq}))$\cite{Kolb}.

At even earlier times, particles of mass $m$ could be 
produced and formed part
of the ``primordial soup'' together with photons when $T> m$.  When the
temperature, that is the
characteristic energy of the photons, was larger than the binding energy of
a certain system, this
system could not survive as such.  This happened with atoms for $T > 1$
eV and with nuclei for
$T > 1$ MeV.  This mean that atoms became
stable for the first
time at the recombination epoch, $T\simeq 1$ eV
 and nuclei were first formed at the BBNS epoch, $T \simeq 10 -0.1$ MeV.
 
   Going to still
earlier times and larger $T$, QCD predicts that chiral symmetry breaking and the
confinement of quarks
within hadrons should  happen at $T \simeq$ 100 MeV.  Even earlier, the
electroweak symmetry should 
be restored at $T > 100$ GeV.  Still earlier, we would encounter physics
beyond the standard
electroweak model, whatever it might be, until the Planck scale $T \simeq
10^{19}$ GeV above which
a quantum theory of gravity would be required.

\subsection*{\bf 5. Relic Abundances of Particles}

 Let us follow the histories of particles that were in equilibrium at a high
temperature $T$ in the primitive Universe.  Stable particles $X$ in
equilibrium have the following  number densities $n_X$ with respect to photons
$n_{\gamma}$:

\noindent
$ - n_X /  n_{\gamma} = f_X (g_X /2),~~~$ for relativistic
particles, $(m_X <<T)$,
where $f_X=1$ for a boson and 3/4 for a fermion, and $g_X$ is the
multiplicity of spin states ($g_X=1$ for a real spin $0$ boson,
$g_X=2$ for the photon and a Weyl or Majorana fermion, $g_X=3$ for a massive
gauge boson, $g_X=4$ for a  Dirac fermion);

\noindent
$ - n_X / n_{\gamma} = (\pi/ 8)^{1/2} (g_X / 2 \zeta(3) )
(m_X / T)^{3/2} exp(-m_X/T),~~~$for non-relativistic particles $(m_X>>T)$.

 Particles go out of equilibrium when their rate of interaction
$\Gamma=n \sigma v$ becomes smaller than the rate at which $T$ decreases,
that is the rate of expansion of the
Universe, the Hubble constant $H$, since $- \dot T/T= \dot a / a =H$
  (or, equivalently, when the mean
interaction time $\Gamma^{-1}$ becomes larger than the age of the Universe
$H^{-1}$).  Particles for which the condition $\Gamma>H$ was never fulfilled
were never in equilibrium and their number has to be computed in other ways
(this is the case of axions, for example).  Different particles go out of
equilibrium at different ``freeze-out" or ``decoupling"
temperatures $T_{f.o.}$, depending on their interactions.

Relativistic particles with weak interactions go out of equilibrium in the
early Universe at $T_{f.o.} \simeq$ 1 MeV.  The estimate of this temperature
is very easy. For light left-handed neutrinos $\nu_L$, 
for temperatures smaller than
the mass of the vector bosons that mediate the interaction, the $W$ and $Z$,
$(M_V \simeq 100$ GeV) but larger than any other masses (of neutrinos,
leptons, quarks ...) the cross section on dimensional grounds
 is $\sigma \simeq G^2_FT^2$, with $G_F$
the Fermi constant, $G_F \simeq  10^{-5}/$GeV$^2 \sim  M_V^{-2}$. 
The  number density of light particles is
$n\simeq T^3$, and their velocity $v\simeq c=1$.  
On the other hand, $H \simeq
{\sqrt G}~ T^2$, with $G$ the gravitational constant, $G \simeq  M_P^{-2}$. 
 Thus the interaction rate goes as 
 $\Gamma \simeq n \sigma v \simeq G_F T^5$, it decreases faster with $T$ than $H$
and at $T= T_{f.o.}$ $\Gamma$ becomes smaller than $H$.  By
equating $\Gamma  = H$ the mentioned $T_{f.o}$ is obtained.

The number $n_{\nu}$ of light neutrinos (that are relativistic at $T \simeq$
1MeV) per co-moving volume is kept constant after their freeze-out, but the
number of photons is increased by the annihilation of $e^+e^-$.  When $T$ drops
below a mass threshold for the production of pairs of certain particles, 
$e^+e^-$ in this case, these
particles cease to be produced and  only annihilate ``heating" the photons and
all the other interacting particles but not the decoupled species. 
Consequently,  the
temperature of the decoupled particles, $T_\nu$ in this case,
 becomes smaller than the photon temperature $T$. 
 Actually, it is not that the temperature of photons increases, 
it just decreases at a smaller rate for a while.
We can compute the ratio $T_\nu /T$ using entropy conservation.
In fact, in thermal equilibrium the total entropy per comoving volume,
 $S = s a^3$, is conserved.
Here $s$ is the entropy density $s = (p+\rho)/T$, thus $s = (4/3) (\rho/T)$
for radiation (and
relativistic particles).  Let us compute the entropy before and after $
e^+e^-$ annihilation and
use entropy  conservation, with the approximation that the scale factor $a$
before and after is the
same, namely assuming there is not enough time for $a$ to change much.
The entropy before  the
annihilation has a contribution due to $e^+e^-$, $s_{e} = 4 \times
(7/8) \times
(2\pi^2/45)T^3$ and  a contribution due to the photons $s_{\gamma} = 2 \times
(2\pi^2/45)T^3$.  After
the annihilation only the photons remain with $T = T_{\rm after} > T_{\rm
before}$, since $(2+7/2)
T_{\rm before}^3 = 2 T_{\rm after}^3$.  The temperature of the neutrinos is
not changed (because neutrinos are decoupled) and remains
equal to $T_{\rm before}$, $ T_\nu = T_{\rm before} = (4/11)^{1/3 }~T$.
Thus, the number of neutrinos plus
anti-neutrinos, $n_{(\nu+{\bar{\nu}})}= n_i$ of every light species $\nu_e,
 \nu_{\mu}, \nu_{\tau}$, that was before $n_i /n_{\gamma}= (3/4)(g_{\nu}/2)$
  becomes $n_i /n_{\gamma}= (3/11)(g_{\nu}/2)$,
   that is, $n_i =(g_{\nu}/2)(115~/cm^3)$, because the number of neutrinos even
   after  their decoupling is  still  proportional to $T^3_\nu$ 
   (see e.g. \cite{Kolb}).

Knowing $T_\nu$ we can compute the contribution of each $\nu$-species that
is still relativistic to
the present radiation energy density $\rho_R$, that is
usually parametrized as $\rho_R = (\pi^2/30) ~g_*
(T)~ T^4$. Here $g_{\ast}(T)= \sum_B g_B (T_B /
T)^4 + {7\over 8} \sum_F g_F (T_F/ T)^4$ is the effective
number of relativistic degrees of freedom at  the temperature $T$, 
the sums run over all relativistic bosons B and 
fermions F at $T$,  $g_B$ and $g_F$ are the number of bosonic and fermionic degrees of freedom and $T_B$ and $T_F$  are the temperatures of each
species, which are equal to $T$ only for species still 
in thermal equilibrium.
 For every $\nu$
species $g_*(T_o) = 2 \times 7/8 \times (T_\nu/T_o)^4 = 2 \times 7/8
(4/11)^{4/3} = 0.454$.  

Assuming all three neutrino species are still relativistic,
photons and neutrinos account for
$g_*(T_o) = 2 + 3 \times
0.454 = 3.36$, and $\Omega_R h^2 = 4 \times 10^{-5} (g_*/3.36)$
(as already  mentioned in section 4).  However,
it is possible
that one or more neutrinos are non-relativistic at present (if $m_\nu > T_o
= 2.3 \times 10^{-4}$ eV), since
the experimental upper bounds are $m_{\nu_e} < 5$ eV, $m_{\nu_\mu} < 160$ keV,
$m_{\nu_\tau} < 24$ MeV.  In this case, the 
contribution of the non-relativistic light neutrinos to the density of the Universe
now is $\rho _{\nu}={\sum_{i=1}^3 }
m_{\nu_i} n_i =\Omega_{\nu}\rho_c$, which means
%
\begin{equation}
\sum_{i=1}^3 m_{{\nu}_{i}} = 92 {\rm eV}~ \Omega_{\nu}
h^2 ({2\over g_{\nu}})~.
\label{mnu}
\end{equation}

\noindent
The sum runs over all non-relativistic
neutrino species lighter than 1 MeV (because we
use $n_\nu$ for neutrinos  relativistic at freeze-out).

From Eq. (\ref{mnu}) an upper bound on neutrino masses follows from an upper bound
on the present density of the Universe, $\Omega_\nu \leq \Omega_o$.
The best bound on $\Omega_oh^2$ comes from a lower bound on the lifetime
of the Universe (see section 3.6).  In a matter
dominated Universe (with
$\Lambda = 0)$, $t_o \gsim 1.3 \times 10^{10}$ yr
 requires $\Omega_oh^2 \lesssim 0.4$, but adding the reasonable bound 
  $h \gsim 0.5$ one obtains 
  $\Omega_oh^2 \lesssim 0.25$. These two bounds imply
 (using Eq. (\ref{mnu}))  $\sum_i
m_{\nu_i} \lesssim 37$ eV and $\sum_i m_{\nu_i} \leq
23$ eV, respectively \cite{Gerstein}. 
 With a non-zero cosmological constant, $\Lambda > 0$, these bounds
are relaxed a bit, since
$\Omega_oh^2$ can be larger for the same $t_o$.
These bounds apply to
the masses $m_{{\nu}_{i}}$ of neutrinos with two degrees of freedom, i.e.
Weyl of  Majorana, and
full weak interactions, thus $g_\nu =2$.  
The right-handed neutrino components, even when they
exist, are not counted,
because either they are never in equilibrium or decoupled much earlier 
than the left handed components, and are therefore much less abundant.

Heavy relics $X$ of mass $m_X$ that become non-relativistic while still in
equilibrium, have a different history.  While annihilations are in
equilibrium, i.e. $\Gamma_{annih.}>H$, the number of non-relativistic particles
decreases with the Boltzmann factor $n_X/T^3 \sim exp(-m_X/T)$.  
When annihilations
cease (because the rate $\Gamma_{annih.}$ becomes small with respect to $H$),
$n_X/T^3$ departs from its equilibrium value and soon it becomes constant.
The larger the
annihilation cross-section is, the longer annihilations remain in equilibrium,
and the smaller is the relic abundance $\Omega$, thus \cite{Lee}
%
\begin{equation}
\Omega h^2 \simeq {1\times 10^{-37}cm^2\over  < \sigma_a v  >}~. 
\label{mX}
\end{equation}
Here $ < \sigma_a v  >$ is the thermal average of the
non relativistic annihilation cross-section  that is always proportional 
to $v^{-1}$, and $<>$
indicates the average over a thermal distribution of momenta (average over
initial states and sum over all final states).

Eq. (\ref{mX}) implies that particles of mass $m_X$ of order GeV- TeV with
annihilation cross sections of  ``weak" order, i.e. $<\sigma_{annih}v > 
\simeq G_F^2 m_X^2
\simeq 10^{-37}$cm$^2(m_X / $GeV$)^2$ for $m_X < M_Z$ and 
$<\sigma_{annih}v > \simeq \alpha/ m_X^2
\simeq 10^{-37}$cm$^2(m_X / $TeV$)^{-2}$ for $m_X >> M_Z$, have 
 relic abundances $\Omega$ of order one,
and may be, therefore, good DM candidates
($\alpha$ is the  electromagnetic constant).

Since elementary particle physicists are exploring extensions of the
electroweak model, plenty of new particles 
 with masses in the few  GeV to a few TeV, are hypothesized.
  Many of them have interactions mediated by particles in the same mass
  range, what means cross sections of the weak
order and, thus,  they can account for
$\Omega \simeq 0.1$ to $1$ and are good DM candidates.
This is the case of the Lightest Supersymmetric Particle, for example.

In the above discussion no asymmetry between $X$-particle and antiparticle 
numbers was
assumed. A cosmic asymmetry  would insure that annihilations  stop when the
minoritary population is depleted, leaving many more
particles than expected without an asymmetry. 

\subsection*{6. Primordial Nucleosynthesis, Also in Crisis?}

When the temperature of the Universe became smaller than the binding energy of
nuclei, at $T \simeq 0.1$ MeV, nuclei  first became stable.  
Because neutrons are heavier
than protons $n_n/n_p$
$\sim{\rm  exp}{[-(m_n-m_p) /T]} < 1$, there are less neutrons than protons.  Most
of the neutrons end up into ${}^4$He (approximately 25\%
of the mass of the Universe) together with an equal number of protons.
 The remaining protons stay as H (approximately 75\% of the
mass of the Universe), and
trace amounts of D, ${}^3$He  and ${}^7$Li  are produced (with 
$n_{{\rm D},{}^3{\rm He}}/n_{\rm H} \simeq O(10^{-4})$ and 
$(n_{{}^7{\rm Li}}/n_{\rm H}) \simeq  O(10^{-10}$).
 Primordial abundances of
the light-elements are difficult to infer from observations and experts
differ on the relative
weight they give to different measurements and methods of inference. 
 While already in 1964 it was known that stars can
only produce less then 5\%
of the existing ${}^4$He, only in the early 70's it was shown that D,
being very weakly-bound, is easily destroyed in stars (mostly into ${}^3$He) 
but  it cannot be produced.

 The predictions of BBNS depend
on the value of the baryon-to-photon ratio $\eta \equiv n_B/ n_\gamma$ 
during the  nucleosynthesis, and it is remarkable 
 considering that  the abundances 
are vastly different, that there
 is a range of $\eta \simeq     O(10^{-10})$  for which
realistic abundances are obtained for all the light element.  The prediction of the ${}^4$He
abundance also depends of
the number of relativistic neutrino species in equilibrium during the BBNS,
$N_\nu$.  The abundance
of ${}^4$He increases with $N_\nu$.  Thus present upper bounds on 
${}^4$He give an upper bound on $N_\nu$ of
about 3. $N_\nu$ parametrizes also any non-standard contribution to the
energy density of the
Universe during the BBNS, so it would be easy to
explain a BBNS prediction of $N_\nu$  larger than 3.  However, a
bound  $N_\nu < 3$ would require at least one of the three known 
neutrinos to contribute less than a relativistic neutrino during BBNS. 
For example, a heavy neutrino
 could decay before the BBNS, i.e. with a lifetime $< 10$ sec. Otherwise,
 one would suspect that there is something wrong in the arguments leading
 to the BBNS bound on $N_\nu$.  Precisely a bound  
 $N_\nu < 2.6$ (95$\%$ C.L.), 
 with a value $N_\nu=2.1 \pm$ 0.3 (1 $\sigma$ error), is the claim 
of a recent paper entitled ``Big Bang Nucleosynthesis in Crisis" \cite{Hata}. 
 One of the main differences in this paper
is a lower  assumed range for D+${}^3$He. Because the abundance of ${}^4$He
increases with $\eta$ while that of D and ${}^3$He decrease with $\eta$ 
(${}^7$Li has
a dip, but increases with $\eta$ for
$\eta \gsim 3 \times 10^{-10})$ a lower  assumed
 range for D+${}^3$He, pushes the acceptable range of $\eta$ in this
paper, to larger values. 
Fig. 3 (taken from \cite{Kolb})
shows the dependence of the different abundance on $\eta$ and of the
${}^4$He abundance on $N_\nu$.
Because the abundance of ${}^4$He 
also increases with $N_\nu$, in order to 
prevent  getting too much ${}^4$He with the larger values of $\eta$
selected by the lower D abundance, 
$N_\nu$ must be smaller.
Given  the dispersion in 
different measurements of primordial abundances (26), to claim, a
``crisis" seems premature.  However one should keep in mind that the
primordial abundances of
${}^4$He, D+${}^3$He and ${}^7$Li leave a narrow range of $\eta$ where
agreement with predictions
for all of them is achieved and when observational uncertainties decrease
further in the near future (mainly in the primordial abundance of D 
\cite{Cardall})
 this narrow range will be further squeezed.

\begin{figure}[b!] 
\vspace{7.5 truecm}
\vspace{10pt}
\caption{Predicted primordial abundances as function of $\eta$ and $N_\nu$.}
\label{fig3}
\end{figure}

Let us briefly see where the different dependences with $\eta$ and $N_\nu$ 
of the various abundances shown in Fig. 3
 come from (for more details see e.g. \cite{Kolb}). 
 Let us start our description at 
$t \simeq 10^{-2 }$ sec, when $T
\simeq 10$ MeV, $\gamma$, $\nu$, $\bar \nu$, $e^+$, $e^-$, $n$ and $p$
 are in equilibrium and  $n_n/n_p \simeq 0.9$ due
to the difference in mass between protons and neutrons
$m_n - m_p = 1.293 $ MeV.  At about $t \simeq 1$ sec,
$T \simeq 1$ MeV, weak scatterings and annihilations go out of equilibrium,
 thus $\nu$'s decouple and the ratio $n_n/n_p$ freezes-out at
  $n_n/n_p\simeq 1/6$ and
 decreases from then on only because neutrons decay (thus the BBNS predictions
depend on the neutron lifetime). 

 The exact temperature at which the weak 
decoupling occurs depends on the value of Hubble parameter $H$. 
The larger H, the sooner the decreasing rate of weak
scatterings and annihilations
becomes equal to H, thus the weak decoupling (or freeze-out) occurs
earlier, consequently the  $n_n/n_p$ ratio at freeze-out is
larger and, because practically all neutrons available when nucleosynthesis
finally happens end up in ${}^4$He, the
predicted abundance of ${}^4$He is also larger.  Since the density of
the Universe $\rho$ increases with the number of neutrino species $N_\nu$, 
H $\sim \sqrt\rho$  and consequently the predicted abundance of
${}^4$He  also increase  with increasing $N_\nu$.  

Even if the temperature at this point, $T \simeq 1$ MeV,
is lower than the binding energy of ${}^4$He, the synthesis of this nucleus
(and consequently, that of
heavier nuclei) cannot happen
because the deuterium D  has a very small binding energy, i.e. it
is constantly destroyed by photons with $E > 2.2$ MeV.
Because there are so many more
photons than baryons, we need to wait until the amount of photons in the
tail of the energy spectrum with $E > 2.2$ MeV becomes smaller than the number 
of D nuclei, namely when $n_\gamma \exp
(-2.2$ MeV$/T)\lesssim n_B$.  When this happens, at 
$T \simeq 2.2 $MeV$/\ell n (\eta^{-1})$ the so called ``D
bottleneck" finishes and ${}^4$He is formed. Because until the end of the
``D-bottleneck" period neutrons
continue decaying, a large value of $\eta$ leads to a shorter time for neutrons
to decay what leads to
larger amounts of ${}^4$He produced.  In order to understand the
duration of this period it is necessary to take into account the
reheating of photons due
to $e^+e^-$ annihilation (see section 5) that happens precisely in this period. 
Due to this effect the temperature decreases
less rapidly and the ``D-bottleneck" finishes at $t \simeq 200$ sec
(without reheating it would finish at $t\simeq $ 7.35 sec 
instead). 

 At this point practically all neutrons end up in
${}^4$He thus $n({}^4$He) $= n_n/2$.  Using $n_n/n_p \simeq 1/7$ we obtain
with a back-of-an-envelope calculation the right amount of ${}^4$He, 
$Y(^4{\rm He})$=0.25 :
%
\begin{equation}
Y(^4{\rm He})
= { 4 n({}^4{\rm He})~~ m_p \over 
 {\left [ 4 n({}^4{\rm He}) + n({\rm H}) \right ] m_p}}
={4({n_n / 2}) \over {4({n_n / 2}) + n_p - n_n}} = 
{2 \left ( {n_n / n_p} \right ) \over {1+ \left ({n_n / n_p} \right )}}~ 
\label{He}
\end{equation}
that is precisely 0.25.
Here $n({}^4{\rm He})$, $n({\rm H})$ are the number densities of ${}^4{\rm He}$
and H nuclei and
$m_p$ stands for the mass of both, protons and neutrons.

Trace amounts of
${}^7$Li  are also produced, but not heavier elements.  The elements with
atomic weight 5 and 8 are not
stable and the gap in mass is then too large for heavier elements to be
formed. This gap is bridged in stars
at higher $T$ though three-body collisions.  There are two different
processes through which
${}^7$Li is produced, that cross in a dip at
 $\eta\simeq 3 \times 10^{-10}$.
 
   In addition, some amount of
D and ${}^3$He are left unburnt, as the reactions that burn them into
${}^4$He freeze-out. The
rates of these reactions are proportional to $\eta$.  So for larger values
of $\eta$ the reactions
that burn D and ${}^3$He are longer in equilibrium and less D and
${}^3$He remain.  

A range in $\eta$ is translated into a range of $\Omega_Bh^2$, since $\eta
= (n_B/n_\gamma) = (\Omega_B ~\rho_c/m_p)/(2 \xi(3)T^3/\pi^2)$. With
$T = 2.73 {}^oK$ one obtains $\eta= 272.2 \times 10^{-10} \Omega_B h^2$.
Bounds on $\eta$ have changed over time, mainly due to observational
 uncertainties and different treatments of the errors. The ranges of the
``Chicago group" have changed from $\eta = (4-7) 10^{-10}$, i.e.  $0.015
\leq \Omega_B h^2 \leq
0.026$ (and $N_\nu < 3.4)$ in 1980's, to $\eta = (2.8-4) 10^{-10}$, i.e.
$ 0.012 \leq \Omega_B h^2
\leq 0.015$ (and $N_\nu < 3.3)$ in 1991, to $\eta = (2-6.5)10^{-10}$,
i.e. $ 0.007 \leq \Omega_Bh^2
\leq 0.024$ (and $N_\nu < 3.9$ at 95\% C.L.) in 1995 \cite{Yang}.  
The newer ``Ohio
group" prefers $\eta =
(3.8-5.2)  10^{-10}$, i.e. $0.014 \leq \Omega_B h^2 \leq 0.019$ (and $N_\nu <
2.6$ at 95\% C.L. as
mentioned above) \cite{Hata}.

As a final comment, let us mention a particularly intriguing explanation for
the result $N_\nu < 3$, if confirmed. 
 The solution may be  an unstable $\nu_\tau$ with
mass in the MeV range
decaying with a lifetime $\tau < 10$ sec into invisible particles 
\cite{Dodelson},  i.e.  $\nu$'s or $\nu\phi$ with
$\phi$ a Goldstone boson, a Majoron.  In particular if the daughter
particles include $\nu_e$ and
$m_{\nu_\tau} = 1-10$ MeV, these extra $\nu_e$ would deplete $n_n$ due to
the process $\nu_e ~n \rightarrow
p~e^-$, because the opposite reaction $\bar\nu _e~p \rightarrow n~ e^+$ requires
an energy of the daughter neutrinos $E_{\bar\nu_e} =
m_{\nu_\tau}/2 > m_n-m_p \simeq $ MeV and is suppressed for $m_{\nu_\tau} <$ 10 MeV.  This
effect can be played off
against the increase in $n_n$ (and consequent increase in ${}^4$He) due to
extra contributions to the density $\rho$ during the BBNS
 to allow even for $N_\nu$ = 16$~$!

\subsection*{7. Structure Formation}

The Universe looks lumpy at scales $\lambda \simeq 100$ Mpc, we see
galaxies, clusters,
superclusters, voids, walls. But it was very smooth at the
surface of last scattering
of the CMBR (i.e. at electromagnetic decoupling, when ions and electrons first
formed atoms) and later
(see section 3). Inhomogeneities have been seen as anisotropies in the
CMBR, so
$(\delta\rho/\rho)$ cannot be much larger than
$\delta T/T \lesssim 10^{-5}$. So
inhomogeneities in density
start small and grow through the Jeans (or gravitational) instability: 
gravitation tends to further empty underdense regions and  to
further increase the density of  overdense
regions.  One can 
follow analytically the evolution due to gravity of the density contrast
$\delta\rho \equiv
(\rho(x) - \rho)/\rho$ (where $\rho$ is the average density) in the linear 
regime where $\delta\rho/\rho < 1$.
In a static fluid the rate of growth of $\delta\rho/\rho$ is exponential, but
in the Universe (an
expanding fluid) it slows down into either a power law, $\delta\rho/\rho
\sim a$, in a matter
dominated Universe or it  stops, $\delta\rho/\rho \simeq$ constant, in
a radiation or a
curvature dominated Universe. We have not so far mentioned
the possibility of a Universe where the curvature term dominates the
  r.h.s. of Eq. (\ref{E1}). Using this equation one can see that 
a  matter dominated open Universe becomes  curvature 
dominated for $a \geq \Omega_o/(1-\Omega_o)$ and no further growth of
the contrast of perturbations in matter can occur. 

Perturbations have different  physical linear dimensions $\lambda$
 which increase with  the Hubble expansion, 
$\lambda = a(t) \lambda_{\rm com}$. Here $a(t)$ is the scale factor  and
$\lambda_{\rm com}$ is the 
linear dimension measured in comoving coordinates (those 
that expand with the Hubble flow). With the usual choice of $a = 1$ 
at present, $\lambda_{\rm {com}}$ is the present actual linear dimension.  
Since $a\sim t^\alpha$ with $\alpha < 1$ (see Eq. (\ref{aH})) while 
the horizon $ct$ grows linearly with $t$, the horizon increases 
with time even in comoving coordinates, encompassing more material as
time goes.  When $\lambda = ct$  we say the
perturbation  of size  $\lambda$ ``enters" into the horizon,
 we could better say that the
perturbation is first  encompassed by the horizon. This moment is called 
``horizon-crossing" and it happens  at
different times for different linear scales $\lambda$, larger scales
 cross later.  
 
 What is the origin of these primordial
fluctuations?  In the standard BB
model perturbations can at most cross the horizon only once,
 coming  from outside the
horizon, where they cannot be generated because
 there can be no causal interactions.
In inflationary models each perturbation  crosses the horizon twice. 
Inhomogeneities in density are generated as quantum
fluctuations, expand exponentially  going outside the
horizon during inflation and re-enter later, after inflation is over.  
Inhomogeneities could also be generated within the horizon
by defects, such as strings and textures, that are remnants
from phase transitions in the early Universe.

  It is convenient to specify the spectrum of
fluctuations at horizon-crossing,
$(\delta\rho/\rho)_{\rm hor}$.  A Harrison-Zel'dovich spectrum is scale invariant  
at horizon-crossing, namely $(\delta\rho/\rho)_{\rm hor}$ = constant.  COBE
observations have shown this spectrum is in fact scale invariant or very 
close to it.  

After horizon-crossing, physical interactions act upon the inhomogeneities and 
generate a ``processed" spectrum.  Three main processes compete.  
Gravitational interactions tend to produce a gravitational collapse,
pressure changes tend to produce pressure waves, and  there may be damping,
caused by the free streaming of non-interacting particles (such as
relativistic neutrinos) or due to collisions (this is the case of the Silk
damping for baryonic matter, due to photon viscosity). 
A gravitational collapse only occurs if the distance a pressure wave travels 
before the collapse,
$v_{\rm sound}$ $t_{\rm free-fall} \equiv \lambda_{\rm Jeans}$, is smaller
than the size $\lambda$ of the inhomogeneity, i.e. $\lambda > \lambda_{\rm Jeans}$.
If so, roughly speaking, pressure waves cannot scape from the collapsing 
inhomogeneity. This is  Jeans criterium for a collapse.  
 
 For baryons before recombination $v_{\rm sound} \simeq
c/\sqrt 3$ (since $v_{\rm
sound}^2 = dp/d\rho$ and the pressure is due to photons, thus $p = \rho/3$),
thus the Jeans length is
larger than the horizon and the density contrast 
$(\delta\rho/\rho)_B$ of perturbations in
baryons within the horizon cannot grow.  After recombination, matter (atoms)
 decouple from radiation (since atoms are neutral), the  pressure is due to
non-relativistic H atoms thus the sound speed and consequently the Jeans
length decrease steeply and perturbations in baryons can
star growing.  Moreover, Silk damping at decoupling erases inhomogeneities in
baryons smaller than those corresponding to clusters of galaxies.  

We can now see that structure could not have been formed
with baryons alone.  Because the Universe has become matter dominated 
at $t_{\rm  eq}$ before recombination,
$t_{\rm  eq}< t_{\rm rec}$, perturbations in baryons can grow as
$(\delta\rho/\rho)_B \sim a$ from recombination onwards.  
However, the maximum growth factor since then is
 $a_o/a_{\rm rec} \approx 10^3$, and with
 $(\delta\rho/\rho)_{\rm rec} < 10^{-4}$, as shown by 
 COBE and other CMBR anisotropy experiments,
$(\delta\rho/\rho)_B$ at present could not have reached 1. 
However, collapsed objects (galaxies,
clusters etc.) have $\delta\rho/\rho > 1$. 
Therefore, we conclude that there is not enough time from recombination
to the present for inhomogeneities in baryons to grow to
$(\delta\rho/\rho)_B \simeq 1$ (after that,
in the ``non-linear" regime, $\delta\rho/\rho$ grows very fast)
 unless baryons after recombination (i.e. when
finally they can collapse into structures) fall into already existing 
potential wells.  These potential wells can only be formed by matter
not coupled to photons, namely dark matter (DM).
 Perturbations in the DM can start growing  before recombination, 
as soon as the Universe becomes matter dominated
(by the DM itself), so they have enough time
to become non linear by now. By recombination the DM has formed
the potential wells into which baryons fall.
The necessity of a matter component in which density perturbations at the
moment of recombination
may be much larger than the perturbations in baryons, is a strong
motivation for the existence of DM.  

The question of which structures are formed first leads
to the distinction of
three types of DM: hot, warm and cold.  The following naive picture of
how fluctuations evolve
helps understanding the differences of these three types.

Let us define as $t_{\rm gal}$, the moment at which the growing horizon
encompasses for the first
time a perturbation of the size of a typical galaxy that contains, $10^{11}$
M$_\odot$ (including the dark
 halo), where M$_\odot$ is the mass of the Sun.  The temperature of the
Universe at that moment
turns out to be $T \simeq 1$ keV.  Hot DM particles are still
relativistic at $t_{\rm gal}$, i.e.
their mass is $m < 1$ keV .  In this case, at $t_{\rm gal}$ each of the
particles in a fluctuation
of the size of the galaxy has moved from its original position a distance
$ct$ equal to the size of
the fluctuation.  The volume occupied by the particles will expand with the
horizon.  Thus, a
moment later the fluctuation is erased. This is damping by free streaming.
At the moment when the  DM
particles become non-relativistic, their motion becomes negligible.  Thus
the smallest structure
that can survive the ``free streaming" of relativistic particles and grow,
is that one encompassed by
the horizon when the particles are becoming non-relativistic.  Light
neutrinos are hot DM and the
smallest structure that can survive contains the Jeans mass,
$M_{\rm Jeans} = 3 \times 10^{15} M_\odot/ (m_\nu/30 eV)^2$
that corresponds to a supercluster. Therefore,
with hot DM (HDM) (such as light neutrinos of $m \lesssim 30 eV$) 
superclusters of galaxies form first
and galaxies must form later, through the fragmentation of the larger
structures.  Simulations have
shown, however, that there is not enough time to form galaxies.  Thus HDM
is rejected as the dominant DM component.  
Particles of mass $m \simeq$ keV, that are just
becoming non-relativistic at
$t_{\rm gal}$, are warm DM.  With warm DM galaxies form first, but
barely.  Cold DM (CDM) is such that perturbations smaller than a galaxy,
 even of the size of the smallest dwarf galaxy
$(10^6$M$_\odot)$, are not erased and can grow.  

Simulations have shown that CDM must be the most abundant form of matter,
 because the``processed" spectrum of perturbations
(i.e. the present spectrum once the effect of physics
acting within the horizon has been considered)  generated in
standard CDM models reproduces the observations
within 10\%. Standard CDM models, make the simplest assumptions, namely 
$\Omega_{\rm CDM} +\Omega_B \simeq \Omega = 1$, $\Lambda =0$,  scale invariant  perturbations at horizon 
crossing, and a scale independent ``biasing" by
which only the highest peaks in the CDM density distribution end up forming 
galaxies.  There is only one feature in the
processed spectrum of CDM perturbations, a change of slope at 
the present scale that corresponds to the horizon at the moment of
 matter-radiation equality, $\lambda _{\rm eq}$.
  Perturbations with $\lambda < \lambda_{\rm eq}$ entered into the horizon
 at $t < t_{\rm eq}$, when the Universe is
radiation dominated.  They cannot grow while the Universe is
radiation dominated, so they all start growing together at $t = t_{\rm eq}$
and they roughly have the same amplitude today if they all started with the
same amplitude at horizon crossing. Perturbations with
$\lambda > \lambda_{\rm eq}$, instead, entered into the horizon at
 $t > t_{\rm eq}$, when the Universe is
matter dominated, so they started growing immediately. Consequently, 
perturbations at larger scales  entered later,
had less time to grow and their amplitude is smaller at present.
  Once $\lambda_{\rm eq}$  (the location of
the change slope) is fixed, the
only remaining free parameter is an overall normalization of the 
CDM spectrum, now provided by the CMBR
anisotropy measured by COBE at large scales, $\theta > 20^o$.  Density
perturbations at these large scales
entered into the horizon very recently (so they did not grow much), thus
providing a measurement of $(\delta\rho/\rho)$ at horizon crossing,
$(\delta\rho/\rho)_{\rm hor}$. 
For more details on this section so far see e.g. \cite{Kolb}.

 While both the shape and normalization so obtained are
almost right, they do not fit the observations \cite{Ostriker}. 
 In  Fig. 4 \cite{Kolb2} the power spectrum of density perturbations 
predicted by standard
CDM models (solid line) normalized by COBE data (the box
on the l.h.s.) and other data (the points with vertical error bars)
are shown. 
As can be seen in the Fig. 4 (solid line) the spectrum 
of standard CDM models has too much
power on small scales (large k$\sim \lambda^{-1}$), the scales of galaxy clusters and smaller.  
Also in Fig. 4 one can see the failure of
HDM to account for the data (see the short-dashed curve) mentioned above.

\begin{figure}[b!] 
\vspace{6.0 truecm}
\vspace{10pt}
\caption{Comparison of the measured power spectrum of density
 perturbations and the predictions of several DM models (notice that
 k$\sim \lambda^{-1}$).}
\label{fig4}
\end{figure}

  Once the normalization given by COBE is fixed, there are
several possibilities to change the spectrum to agree with observations.
Because HDM tends to erase  structure at small scales,
one of the solutions consists in adding to the CDM a bit of HDM, namely
neutrinos,  in what are called  mixed DM (MDM) or hot-cold DM (HCDM) 
models \cite{Shafi}. See the dotted line in Fig. 4. 
In particular, models with $\Omega_\nu = 0.2$, what corresponds to
 $\sum_i {m_{{\nu}_{i}}}$ = 5 eV, and the rest of $\Omega_o$ in CDM plus
some baryons, with $\Omega = 1$ and  $\Lambda = 0$ works well. 
Another possibility is that of a ``tilted''
primordial spectrum of fluctuations at horizon crossing, one that slightly 
favors larger scales over smaller scales
(instead of the flat, scale invariant, Harrison-Zel'dovich spectrum) within
the COBE observational limits.  This is called ``tilted'' CDM (TCDM) \cite{Cen}. 
See the long-dashed line (labelled with n=0.8) in Fig. 4. 

 Another family of solutions is obtained by realizing that
a shift towards larger scales of the only feature in the CDM spectrum, i.e.
$\lambda_{\rm eq}$, the scale where the
slope changes, is enough to provide good agreement with observations, since
it effectively amounts to increasing the power of the spectrum at scales 
larger than the break point
$(\lambda > \lambda_{\rm eq})$ with respect to those smaller  than it
$(\lambda < \lambda_{\rm eq})$.
The break in the slope occurs at the moment of radiation-matter equality,
$\rho_{\rm rad} = \rho_{\rm matter}$, when the Universe becomes matter
dominated (by the DM).  The scale of the break,
$\lambda_{\rm eq}$, corresponds to the present size of the horizon at
matter radiation equality, $ct_{\rm eq}$. It is 
$\lambda_{\rm eq} \simeq 10 $Mpc$ \left [ \left ({g_* / 3.36}\right
)^{1/2}  (\Omega_o h^2)^{-1}
 \right  ]$,
where $\Omega_o$ is essentially the present matter density,
since $\Omega_R << \Omega_o$
(recall that the total density is $\Omega = \Omega_o +
\Omega_{\rm vac}$, if the cosmological constant $\Lambda \not= 0$) and
$g_*$ is the effective number of relativistic degrees of freedom at 
$t_{\rm eq}$, $g_* = 3.36$ for photons plus three relativistic neutrino 
species (see section 5).  

The equation $\lambda_{\rm eq} \equiv {(10 h^{-1}Mpc)  \Gamma^{-1}}$ 
defines the ``shape
parameter'' \cite{Efstathiou} $\Gamma \equiv \Omega_o h (g_*/3.36)^{-1/2}$ 
  (notice that distances determined through red-shifts
 are given in $h^{-1} Mpc$ units (see Eq. (5)).
The data require $\Gamma\simeq 0.25 \pm 0.05$, while standard
CDM models (with the standard choices of  $h = 0.5$,
$\Omega_o = 1$, $g_* = 3.36$), has $\Gamma =0.5$.  In fact, as we have
explained a larger $\lambda_{\rm eq}$, thus a smaller $\Gamma$,
 would provide agreement with data.  In order to  lower the value
of $\Gamma$ with respect to that of the standard CDM model one needs
to either:  1- lower $h$ \cite{Bartlett} (what implies an older Universe),
or 2- increase $g_*$
(namely increase the radiation content of the Universe at $t_{\rm eq}$), 
 or 3- lower $\Omega_o$
(i.e. take $\Omega_o < 1$), so that we either live in an open Universe
(open CDM models, OCDM) if $\Lambda = 0$, or in a Universe
with a cosmological constant that provides $\Omega_{\rm vac} = 1 -\Omega_o$ 
 ($\Lambda$CDM models \cite{Turner}), 
for example with $h = 1$, $\Omega_{\rm CDM} = 0.18$, $\Omega_B = 0.02$,
$\Omega_{\rm vac} = 0.80$,   or 4- a combination of all three above. 
We have already examine the consequences of the 3rd. solution in relation
with the possible ``age crisis", in section
 3.6. Let us examine the consequences of the other two.

 If we want to keep $\Omega_o = 1$, with the standard value of
 $g_* = 3.36$  one would need $h = 0.3$ to get $\Gamma
= 0.3$, but this would
lead to a very old Universe with $t_o = 2.2 \times 10^{10}$yr. 
 If instead we keep the standard value of $h$  i.e. 
$h = 0.5$ (this low-value of $h$ is necessary with 
$\Omega_o = 1$ to account for $t_o = 1.3 \times 10^{10}$yr),
the value $g_* = 9.33$ is necessary to get $\Gamma = 0.3$, what amounts to
the equivalent of 16 relativistic neutrino species.  However, primordial 
nucleosynthesis does not allow for much more
than 3 neutrino species (except with a heavy unstable $\nu_{\tau}$, 
see section 6 and below).  

A way to obtain the large amount of radiation needed is through a heavy
neutrino decaying into radiation, with the right combination of mass and
lifetime, in so-called $\tau$CDM models \cite{Bardeen}\cite{Dodel}.
A massive neutrino matter dominates the energy density of the Universe as soon
as it becomes non-relativistic $m_\nu \gsim T$
(since $n_\nu \simeq n_\gamma$ and $\rho_\nu
= n_\nu m_\nu$, $\rho_{\rm rad} \simeq n_\gamma T)$, thus their decay 
products radiation-dominate the Universe at decay.  
For $m_\nu < 1$MeV the right mass-lifetime combination lie on a
narrow strip around the (usually called) ``galaxy formation'' bound.  
This bound \cite{Hut} is obtained by requiring that the decay
products of a massive neutrino cease to dominate the energy density of the
Universe before baryons
should start falling into gravitational wells, at recombination 
$t_{rec}\simeq 10^{-5}t_o$, what yields
$\tau \leq (92 {\rm eV}/m_\nu)^2 (\Omega_oh^2)^2 10^{-5}t_o$.  Near this bound,
at the boundary between being irrelevant and harmful, unstable neutrinos
 could help in the formation of structure in the Universe.  A
heavier neutrino, of $m_\nu \simeq 1 - 10$ MeV, necessarily $\nu_\tau$,
decaying at or just before nucleosynthesis, $\tau = 0.1 - 100$ sec, 
would also provide a solution \cite{Dodel}.  If $\tau $= 0.1-few sec and
$\nu_e$ are among the decay products,
 neutrons would be depleted, 
leaving room for the presence of
 up to 16 equivalent  relativistic neutrino species during nucleosynthesis
 \cite{Dodelson},
 as explained at the end of section 6.
This $\nu_\tau$ decays too fast to get to matter dominate so the decay 
products never dominate the energy density of the Universe and  additional  
unknown radiation has to be added ad-hoc to increase $g_*$ as needed.  
If instead $\tau_{\nu}\simeq 10 - 100$ sec, the $\nu_\tau$ can
dominate the energy density of the Universe  before decaying and its decay
products provide the necessary extra radiation.  
The $\nu_\tau$ decay modes  involved here should all be into neutral
particles, $\nu_\tau \rightarrow 3~\nu'$s or $\nu_\tau \rightarrow \nu \phi$,
 with $\phi$ a Majoron (a zero mass Goldstone boson) for example.  
All visible modes, i.e. producing electrons or photons, are forbidden in the
necessary range.

All these modified CDM models seem to be able to fit present data, 
however they predict very different patterns of
acoustic peaks in the CMBR
anisotropy power spectrum $\ell(\ell+1)C_\ell$ (see section 3.3 
and Eq. (\ref{deltaT})). The location and relative height of the peaks between
$\ell\simeq 200$ and  several 1000 characterize different models.  The detection of the
CMBR anisotropies so far obtained with the COBE satellite and balloon 
experiments, does not discriminate between models,
but the next generation of observations will do it.  There are two approved
new satellite experiment with sub-degree resolution, 
MAP in the U.S. and  COBRAS/SAMBA in Europe.  They will give results as shown 
in the  Fig. 2  \cite{Scott}
together with  the predictions
of four different models, three of them mentioned above SCDM (standard CDM),
 $\Lambda$CDM (CDM with $\Lambda >$ 0) and
OCDM (open CDM), and one labelled with ``strings'', that corresponds to
structure formation due to inhomogeneities caused in a phase transition 
that produced ``cosmic strings'' (vortex-like solitonic remnants).  
The dark band around the SCDM line shows the expected experimental
error in the $C_\ell$'s for an experiment like COBRAS/SAMBA, 
with $10'$ beam-size and 10 $\mu K$ noise (that would
allow to resolve
anisotropies up to $\sim 0.5^o $ for $\Omega = 1)$  \cite{Scott}.  It is clear from 
 Fig. 2
that the different theoretical models  shown could be easily distinguished.

\subsection*{8.  Detection of the Dark Matter in the Halo of Our Galaxy}

Our own galaxy is spiral.  The Sun is part of a thin disk of about 12 kpc
of radius consisting
mainly of stars and gas clouds orbiting the galactic center.  The Sun is at
$r_o \simeq 8.5$ kpc from the center and moving at about 
$v_c\simeq$ 220 km/sec. The
rotation curve of the Milky Way
is flat at about $v_c$ out to as far as it is measured, showing that our
galaxy has a spherical or
ellipsoidal dark halo.  Assuming a spherical and isotropic velocity
distribution, the density
profile of this halo is usually parametrized as $\rho(r) = \rho_o \left
[({a^2+r_o^2)/ (a^2+r^2)}\right ]$ where
 $\rho_o \simeq 0.01 M_\odot/$pc$^3 \simeq 0.4$ GeV/cm$^3$ is the 
halo density in the solar neighbourhood and $a$ is the core radius of the halo
$(\rho\simeq$ const at $r < a)$,
that has been estimated at $a \simeq 2 -10$ kpc.  There is also a central
luminous component of the
galaxy, the spheroid or bulge.  The distribution and motion of the luminous
matter trace the
combined gravitational effects of the three components, the disk, the bulge
and the halo.  The halo dominates at large radius.

In the last ten years, attempts have been made to detect the galactic DM in
experiments that could
reveal its nature.  Because the DM in galactic haloes amount to
$\Omega_{\rm haloes} \simeq
0.02h^{-1}$ (see section 3.7) nucleosynthesis bounds on the baryonic 
density $\Omega_B$ (see section 6) do not preclude the
possibility that ``dark'' baryons constitute some or all the DM in galaxy
haloes like ours.
Candidates for dark baryons could be macroscopic objects up to the size of
``brown dwarfs''. These are stars
too small, $m < 0.08 M_\odot$, for their centers to reach  a sufficiently high
temperature to burn
hydrogen (as in a normal star).  These candidates are called MACHOs (for
massive compact halo
objects) as opposed to WIMPs (for weakly interacting massive particles), that
 are elementary particle
candidates for the DM.  Elementary particle candidates include also
massive neutrinos (actually
the only DM particle candidates known to exist, 
but not known to be massive as needed).
The three known neutrinos, if stable, are hot DM candidates, and the energy 
they would carry is too small to allow their
detection  in the  dark halo of our galaxy 
(but the measurement of their masses in
the range 1-few 10 eV
could tell).  The theoretically preferred cold DM candidates are WIMPs,
with masses in the range 1 GeV- 1 TeV, and the axion,
 a very light boson that appears in the Peccei-Quinn solution to the
strong CP problem of elementary particles, with a mass $m \simeq
10^{-5} $eV. 

 WIMPs appear in many extensions of the standard model of 
elementary particles, such as the
minimal supersymmetric extension of the standard model, where the usual
WIMP candidate is the lightest ``neutralino'' (the lightest fermionic 
partner of  the  neutral gauge and Higgs bosons).  As we have
seen (see Eq. (\ref{mX})), it is enough for a particle in the GeV-TeV mass range to 
have annihilation cross sections of
weak order in order to have a relic abundance $\Omega\simeq 0(1)$. 
 WIMPs are being search for in direct
and indirect DM searches
and accelerator experiments.  If these particles compose the halo of our
galaxy they may be
detected directly, through the energy they may deposit in collisions with
nuclei within detectors,
or indirectly through their annihilation products, coming from
annihilations in the Sun or Earth,
where they may accumulate, or in the halo itself.  Direct WIMP searches look
for energies of order
$mv^2\simeq 0$(10 keV) deposited in ionization and phonons, in kg-size
cryogenic crystals, or
producing scintillation or excitations of superfluids and superconductors.
Indirect WIMP detectors
search mainly for high energy neutrinos produced in WIMP annihilations in
the Sun and the Earth, in
large water or ice Cerenkov light detectors (such as Superkamiokande,
AMANDA, DUMAND and NESTOR) or
underground detectors (such as MACRO). For a review of particle
DM searches see e.g. \cite{Sadoulet}.  

Axions could be detected through
their coupling to two
photons, looking for the resonant conversion of halo axions to microwave
photons in a cavity with a
magnetic field.  The first detector of this type capable to actually test
halo axions is based at
Livermore \cite{Bibber} and has started taking data recently.

The most fruitful DM searches so far, have been those looking for MACHOs
though the gravitational microlensing \cite{Paczynski}
of background stars in the Large Magellanic Cloud (LMC), an
irregular galaxy
satellite of the Milky Way.   
In a gravitational lensing event an intervening object located in the line 
of sight of a luminous object deflects the light coming from it,
producing multiple images.  In a microlensing event, these multiple images
are superposed, what
is seen as an increase in the brightness of the lensed object.  Three
independent groups the
American-Australian MACHO, the French EROS and the Polish-American OGLE
collaborations (that seem
to be competing for the most politically incorrect name) have seen
microlensing events.  They were
looking for (and found in 1993 \cite{Macho}) events in
which a star from the LMC or the bulge of
our galaxy symmetrically
brightens and fades, as a MACHO (or a not seen faint star) passes near its 
line of sight (one could describe the event
 as a symmetric ``anti-eclipse'') during a time $\Delta t$ that depends
on the lensing mass $M$,
$\Delta t = 140$days$(M/M_\odot)^{1/2}$.  So far about 90 events have been
 seen in the direction of the central bulge of our galaxy (where most 
lenses are expected to be faint stars in the bulge itself)
and about 10 towards the LMC.  These numbers
are about a factor of 5 too small from what was expected from
the LMC if the dark halo of the Milky Way consisted of MACHOs
and about a factor of 3 too large from what was expected from 
the bulge (see e.g. \cite{Bennett}). 
 
The number of events towards the LMC is too small to
be compatible with a dark halo dominated by MACHOs (within the mass range
tested $M\simeq  10^{-7} -1 M_\odot)$. Machos could account for up to 
 $\sim 30\%$ of the expected 	halo mass, unless the halo
is much lighter than presently believed \cite{gates}.  The large number of microlensing
events towards the
bulge, where the lensing objects are expected to be mostly faint stars, is
much larger than
expected on the basis of assuming a spherical bulge in the center of our
galaxy, as seen in regular
disk galaxies.  The excess indicates that the bulge of
our galaxy is a bar with its axis oriented towards the Sun \cite{Zhao}, 
in agreement with other observations,
 what would
yield a larger star density
along our line of sight towards the bar.  An alternative explanation could
be that the mass of the
disk is much larger than in standard galactic models.  More data will allow
to distinguish these
possibilities.

\subsection*{Acknowledgments}
I thank the organizers of this workshop for their invitation.
This work was supported in part by the U.S. Department of Energy under Grant
DE-FG03-91ER 40662 TaskC.

\end{document}